\documentclass[letterpaper]{article} 
\usepackage{aaai25}  
\usepackage{times}  
\usepackage{helvet}  
\usepackage{courier}  
\usepackage[hyphens]{url}  
\usepackage{graphicx} 
\usepackage{natbib}  
\usepackage{caption} 
\usepackage{algorithm}
\usepackage{algorithmic}
\usepackage{amsmath}
\usepackage{multirow}
\usepackage{microtype}
\usepackage{amssymb}
\usepackage{xspace,mfirstuc,tabulary}
\usepackage{tabularx}
\usepackage{booktabs}
\usepackage{newfloat}
\usepackage{listings}
\DeclareCaptionStyle{ruled}{labelfont=normalfont,labelsep=colon,strut=off} 
\floatstyle{ruled}
\newfloat{listing}{tb}{lst}{}
\floatname{listing}{Listing}
\title{Logic Induced High-Order Reasoning Network for Event-Event Relation Extraction}
\author{
    Peixin Huang\textsuperscript{\rm 1}, Xiang Zhao\textsuperscript{\rm 2}, Minghao Hu\textsuperscript{\rm 3}, Zhen Tan\textsuperscript{\rm 1}\thanks{Corresponding author.}, Weidong Xiao\textsuperscript{\rm 1}
}
\affiliations{
    \textsuperscript{\rm 1}National Key Laboratory of Information Systems Engineering, National University of Defense Technology, China\\
    \textsuperscript{\rm 2}Laboratory for Big Data and Decision, National University of Defense Technology, China\\
    \textsuperscript{\rm 3}Information Research Center of Military Science, China

    \{huangpeixin15, xiangzhao, tanzhen08a, wdxiao\}@nudt.edu.cn, huminghao16@gmail.com
}

\usepackage{bibentry}

\begin{document}

\maketitle

\begin{abstract}
To understand a document with multiple events, event-event relation extraction (ERE) emerges as a crucial task, aiming to discern how natural events temporally or structurally associate with each other. To achieve this goal, our work addresses the problems of temporal event relation extraction (TRE) and subevent relation extraction (SRE). 
The latest methods for such problems have commonly built document-level event graphs for global reasoning across sentences. However, the edges between events are usually derived from external tools heuristically, which are not always reliable and may introduce noise. Moreover, they are not capable of preserving logical constraints among event relations, e.g., coreference constraint, symmetry constraint and conjunction constraint. These constraints guarantee coherence between different relation types, enabling the generation of a unified event evolution graph. 
In this work, we propose a novel method named LogicERE, which performs high-order event relation reasoning through modeling logic constraints. Specifically, different from conventional event graphs, we design a logic constraint induced graph (LCG) without any external tools. LCG involves event nodes where the interactions among them can model the coreference constraint, and event pairs nodes where the interactions among them can retain the symmetry constraint and conjunction constraint. 
Then we perform high-order reasoning on LCG with relational graph transformer to obtain enhanced event and event pair embeddings. Finally, we further incorporate logic constraint information via a joint logic learning module.
Extensive experiments demonstrate the effectiveness of the proposed method with state-of-the-art performance on benchmark datasets.
\end{abstract}

\section{Introduction}
\begin{figure}[h]
	\centering
	\includegraphics[width=0.8\linewidth]{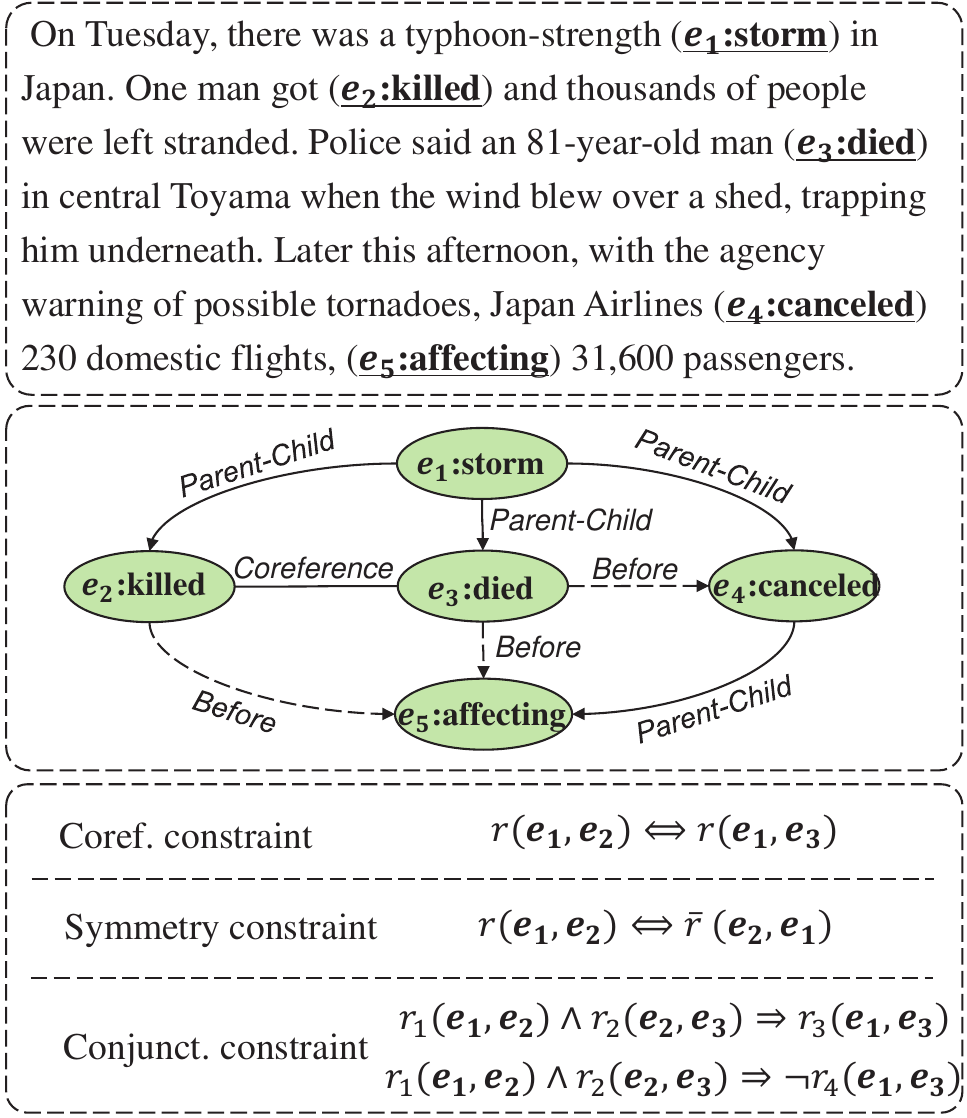}
	\caption{An example of an event evolution graph described in
		the document.}
	\label{samples}
\end{figure}

Interpreting news messages involves identifying how natural events temporally or structurally associate with each other from news documents, i.e., extracting event temporal relations and subevent relations. Through this process, one can induce event evolution graphs that arrange multiple-granularity events with temporal relations and subevent relations interacting among them. The event evolution graphs built through event-event relation extraction (ERE) are important aids for future event forecasting~\cite{DBLP:conf/emnlp/ChaturvediPR17} and hot event tracking~\cite{DBLP:journals/ipm/ZhuangFH23}.
As shown in Figure~\ref{samples}, ERE aims to induce such an event evolution graph, in which the event mention \textit{storm} involves more fine-grained subevent mentions, i.e., \textit{killed}, \textit{died} and \textit{canceled}. Some of those mentions follow temporal order, e.g., \textit{died} happens BEFORE \textit{canceled}. Generally, predicting the relations between diverse events within the same document, such that these predictions are coherent and consistent with the document, is a challenging task~\cite{DBLP:journals/access/XiangW19}.

Recently, significant research efforts have been devoted to several ERE tasks, such as event temporal relation extraction (TRE)~\cite{DBLP:conf/aaai/ZhouYHCCSP021,DBLP:conf/coling/WangL022,DBLP:conf/eacl/TanPH23} and subevent relation extraction (SRE)~\cite{DBLP:conf/aaai/ManNVN22,DBLP:conf/acl/HwangLYPZM22}. Nonetheless,
ERE is still challenging because most event relations lack explicit clue words such as \textit{before} and \textit{contain} in natural languages, especially when events scatter in a document. Accordingly, a few previous methods attempt to build a document-level event graph to assist in the cross-sentence inference, where the nodes are events, and the edges are designed with linguistic/discourse relations of event pairs~\cite{DBLP:conf/dasfaa/ZhuangHZ23,DBLP:journals/corr/abs-2403-12523}. Despite the success, these methods face two major issues. First, they are not capable of preserving logic constraints among relations, such as transitivity, during training time~\cite{DBLP:conf/conll/RothY04}. Second, the edges heuristically from external tools may introduce noise and cause exhaustive extraction~\cite{DBLP:conf/naacl/PhuN21}. 

For the first issue, \citet{DBLP:conf/emnlp/WangCZR20} propose a constrained learning framework which enforces logic coherence amongst the predicted relation types through extra differentiable objectives. However, since the coherence is enforced in a soft manner, there is still room for improving coherent predictions. In this work, we show that it is possible to enforce coherence in a much stronger manner by feeding logic constraints into the event graphs.
For the second issue, \citet{DBLP:conf/coling/00010DLWSZ22} propose an event pair centered causality identification model which takes event pairs as nodes and relations of event relations as edges, refraining the external tools and enabling the causal transitivity reasoning. However, some useful prior relation constraints such as coreference are discarded. Moreover, we observe logical property information loss from the document to graph, which should be complied in global inference for ERE. 
We summarize these properties implied in the temporal and subevent relations as three logical constraints: (1) \textbf{Coreference constraint}: Considering the example in Figure~\ref{samples}, given that \textit{$e_2$:killed} is BEFORE \textit{$e_5$:affecting} and \textit{$e_3$:died} is COREFERENCED to \textit{$e_2$:killed}, \textit{$e_3$:died} should be BEFORE \textit{$e_5$:affecting}. (2) \textbf{Symmetry constraint}: As shown in Figure~\ref{samples}, \textit{$e_1$:storm} is a PARENT of \textit{$e_2$:killed}, indicting \textit{$e_2$:killed} is a CHILD of \textit{$e_1$:storm}. (3) \textbf{Conjunction constraint}: From Figure~\ref{samples}, if \textit{$e_1$:storm} is a PARENT of \textit{$e_3$:died} and \textit{$e_3$:died} is BEFORE \textit{$e_4$:canceled},  the learning process should enforce \textit{$e_1$:storm} is a PARENT of \textit{$e_4$:canceled} by considering the conjunctive logic on both temporal and subevent relations. These logical constraints depict the mutuality among event relations of TRE and SRE, enabling the generation of a unified event evolution graph. 
While previous researches center on preserving these properties with post-learning inference or differentiable loss functions~\cite{DBLP:conf/emnlp/WangCZR20}, there is no effective way to endow the graph model with these logical constraints for global reasoning. 

In this paper, we consider the above constraints and propose a novel ERE model, logic induced high-order reasoning network (LogicERE)~\footnote{After publication, the corresponding AAAI link for our paper will be updated here.}. Our intuition is to feed the logic constraints into the event graphs for high-order event relation reasoning and prediction. Specifically, we first build a logic constraint induced graph (LCG) that models these logical constraints through interactions between events and among event pairs. Then, we encode the heterogeneous LCG with relational graph transformer. This enables the model to effectively reasoning over remote event pairs while maintaining inherent logic properties of event relations. Finally, we introduce a joint logic learning module and design two logic constraint learning objectives to further regularize the model towards consistency on logic constraints. 

LogicERE models these high-order logical constraints in two aspects. Firstly, our proposed LCG consolidates both event centered and event pair centered graphs, so that it can reason over not only coreference property among events, but also high-order symmetry and conjunction properties among event pairs. Specifically, LCG defines two types of nodes, i.e., event nodes and event pair nodes. Accordingly, there are three types of edges: (1) Event-event edge for prior event relations (e.g., coreference), which retains the coreference constraint and model the information flow among event nodes. (2) Event pair-event pair edge for two event pairs sharing at least one event, which keeps the symmetry and conjunction constraints, as well as captures the interactions among event pair nodes.
(3) Event-event pair edge for an event pair and its corresponding two events, which models the information flow from event nodes to event pair nodes. As LCG preserves these inherent properties of event relations, we can get enhanced event and event pair embeddings through high-order reasoning over it.
Secondly, inspired by the logic-driven framework of \citet{DBLP:conf/emnlp/LiGMS19}, we soften the logical properties through differentiable functions so as to incorporate them into multi-task learning objectives. The joint logic learning module enforces our model towards consistency with logic constraints across both TRE and SRE tasks. It is also a natural way to combine the supervision signals coming from two different tasks.

\section{Related Work}
\subsection{Temporal relation extraction} 
Early studies usually employ statistical models combined with handcrafted features to extract temporal relations~\cite{DBLP:conf/acl/YoshikawaRAM09, DBLP:conf/acl/ManiVWLP06}. These methods are of high computational complexity and are difficult to transfer to other types of relations. 

Recently, with the rise of pre-trained language models, various new strategies have been applied to TRE task. To effectively model the long texts, some studies incorporate syntax information such as semantic trees or abstract meaning representation (AMR) to capture remote dependencies~\cite{ DBLP:conf/ictai/VenkatachalamMB21,DBLP:conf/naacl/ZhangNH22}. Others construct global event graphs to enable the information flow among long-range events~\cite{DBLP:conf/ijcai/LiuXCZ21,DBLP:conf/icml/0001WRZ22}. For example, \citet{DBLP:conf/naacl/ZhangNH22} design a syntax-guided graph transformer to explore the temporal clues.  \citet{DBLP:conf/ijcai/LiuXCZ21} build uncertainty-guided graphs to order the temporal events. To deepen the models’ understanding of events, some studies propose to induce external knowledge~\cite{DBLP:conf/emnlp/HanZP20, DBLP:conf/eacl/TanPH23}. \citet{DBLP:conf/emnlp/HanZP20} propose an end-to-end neural model that incorporates domain knowledge from the TimeBank corpus~\cite{2003The}. Some multi-task strategies are also employed for this task ~\cite{DBLP:conf/acl/Huang0ZFL023, DBLP:journals/jamia/KnezZ24}. These strategies can facilitate the models’ learning of complementary information from other tasks. 

\subsection{Subevent relation extraction} 
In earlier studies, machine learning algorithms are utilized to identify the internal structure of events~\cite{DBLP:conf/ijcai/0001LWLJL22, DBLP:conf/lrec/GlavasSMK14}. Afterwards, the introduction of deep learning has led to new advances of this task. \citet{DBLP:conf/acl/ZhouNKR20} adopt the multi-task learning strategy and utilize duration prediction as auxiliary task. \citet{DBLP:conf/aaai/ManNVN22} argue that some context sentences in documents can facilitate the recognition of subevent relations. Thus, they adopt the reinforcement learning algorithm to select informative sentences from documents to provide supplementary information for this task. \citet{DBLP:conf/acl/HwangLYPZM22} enforce constraint strategies into probabilistic box embedding to maintain the unique properties of subevent relations.

To sum up, existing work only focuses on either TRE or SRE, only a few studies seek to resolve both two tasks. \citet{DBLP:journals/ipm/ZhuangFH23, DBLP:conf/dasfaa/ZhuangHZ23}  
adopt the dependency parser and hard pruning strategies to acquire syntactic dependency graphs that are task-aware. \citet{DBLP:journals/inffus/ZhuangFH23} propose to use knowledge from event ontologies as additional prompts to compensate for the lack of event-related knowledge. \citet{DBLP:journals/corr/abs-2403-12523} introduce the features of event
argument and structure to obtain graph-enhanced event embeddings. 
However, the above works regard event relation extraction as a multi-class classification task, and do not guarantee any coherence between different relation types, such as symmetry and transitivity. Different from these works, our LogicERE guarantee coherence through a high-order reasoning graph which is embedded with three essential logical constraints, as well as joint logic learning.

\section{Model}
\label{ssec:Methodology}
This paper focuses on the task of event-event relation extraction (ERE). 
The input document $\mathcal{D}$ is represented as a sequence of $n$ tokens $\mathcal{D}=[x_1, x_2, ..., x_n]$. Each document contains a set of annotated event triggers (the most representative tokens for each event) $\mathcal{E_D}=\{e_1, e_2, ..., e_k\}$. The goal of ERE is to extract the multi-faceted event relations from the document. Particularly, we focus on two types of relations, i.e., Temporal and Subevent, corresponding to the label sets $\mathcal{R}_{Temp}$ which contains BEFORE, AFTER, EQUAL, VAGUE, and $\mathcal{R}_{Sub}$ which contains PARENT-CHILD, CHILD-PARENT, COREF, NOREL respectively. Note that each event pair is annotated with one relation type from either $\mathcal{R}_{Temp}$ or $\mathcal{R}_{Sub}$, as the labels within two sets are mutually exclusive.

There are four major parts in our LogicERE model: (1) Sequence Encoder, which encodes event context in input document, (2) Logic Constraint Induced Graph, which build a event graph preserving logic constraints, (3) High-Order Reasoning Network on LCG, which performs high-order reasoning with relational graph transformer to obtain enhanced event and event pair embeddings, and (4) Joint Logic Learning, which further incorporates logic properties through well-designed learning objectives.

\subsection{Sequence Encoder}
To obtain the event representations and contextualized embeddings of the input document $\mathcal{D}=[x_t]_{t=1}^n$ (can be of any length $n$), we leverage pre-trained RoBERTa~\cite{DBLP:journals/corr/abs-1907-11692} as a base encoder. We add special tokens "\texttt{[CLS]}" and "\texttt{[SEP]}" at the start and end of $\mathcal{D}$, and insert "\texttt{<t>}" and "\texttt{</t>}" at the start and end of all the events to mark event positions~\cite{DBLP:conf/coling/00010DLWSZ22}. Thus, we have:
\begin{equation}
h_1, h_2, ..., h_{n^{'}} = \mathrm{Encoder} ([x_1, x_2, ..., x_{n^{'}}])
\end{equation}
where $h_i \in \mathbb{R}^d$ is the embedding of token $x_i$. We employ the embeddings of token "\texttt{[CLS]}" and "\texttt{<t>}" to represent the document and the events respectively. If the document’s length exceeds the limits of RoBERTa, we adopt the dynamic window mechanism to segment $\mathcal{D}$ into several overlapping spans with specific step size, and input them into the encoder separately. 
Then, we average all the embeddings of "\texttt{[CLS]}" and "\texttt{<t>}" of different spans to obtain the document embedding $h_{\mathtt{[CLS]}} \in  \mathbb{R}^d$ or each event embedding $h_{e_i}\in  \mathbb{R}^d$, respectively.

\subsection{Logic Constraint Induced Graph}
Considering the logic constraints of event relation is based on the motivation that  such logic properties comprehensively define the varied interactions among those events and relations. In this section, we construct a logic constraint induced graph (LCG) which can preserve these logic properties through unifying both event centered and event pair centered graphs.  Specifically, given all the events of document $\mathcal{D}$, LCG is formulated as $\mathcal{G}=\{\mathcal{V}, \mathcal{C}\}$, where $\mathcal{V}$ represents the nodes and $\mathcal{C}$ represents the edges in the graph. We highlight the following differences of $\mathcal{G}$ from previous event graphs and event pair graphs. 

First, there are two types of nodes in $\mathcal{V}$, i.e., the event nodes $\mathcal{V}_e$ and the event pair nodes $\mathcal{V}_{ep}$. Each node in $\mathcal{V}_{ep}$ refers to a different pair of events from $\mathcal{D}$. Instead of merely using events or event pairs as nodes, LCG preserves both of them, enabling high-order interactions through edges.

Second, for edges $\mathcal{C}$, instead of using all edges between any two nodes, we design three types of edges following the logic constraints: (1) Event-event edges $\mathcal{C}_{ee}$ for two events that are co-referenced 
, which is motivated by the \textbf{coreference constraint} in Introduction. These edges are optional. $\mathcal{C}_{ee}$ contributes to event relation reasoning as  co-referenced events are expected to share the same relations with other events. Meanwhile, no additional relations exist between co-referenced events. (2) Event pair-event pair edges $\mathcal{C}_{pp}$ for two event pairs that share at least one event
,which is motivated by the \textbf{symmetry constraint} and \textbf{conjunction constraint} in Introduction. Particularly, for the TRE task, symmetry constraint exists in a pair of reciprocal relations BEFORE and AFTER, as well as two reflexive ones EQUAL and VAGUE. Similarly, the SRE task includes reciprocal relations PARENT-CHILD and CHILD-PARENT as well as reflexive ones COREF and NOREL. The conjunction constraint enables the relation transitivity in a single task, and unifies the ordered nature of TRE and the topological nature of SRE~\cite{DBLP:conf/emnlp/WangCZR20}. (3) Event-event pair edges $\mathcal{C}_{ep}$ for an event pair and its corresponding events.
We design $\mathcal{C}_{ep}$ to bridge the information flow between events and event pairs. 

\subsection{High-Order Reasoning Network on LCG}
We perform high-order reasoning on  LCG, which takes the relation heterogeneity into account and captures diversified high-order interactions within events and event pairs. 

\textbf{Initial Node Embeddings.} \ \ For global inference, we firstly initialize node embeddings. Formally, for the event node $e_i \in \mathcal{V}_e$, we take the contextualized event embeddings from the sequence encoder for initialization:
\begin{equation}
v_{e_i}^{(0)}= h_{e_i} \textbf{W}_n
\end{equation}
where $0$ indicates the initial state and $\textbf{W}_n \in \mathbb{R}^{d\times2d}$ is a learnable weight matrix. 

For the event pair node $e_{i,j} \in \mathcal{V}_{ep}$, we concatenate two corresponding event embeddings:
\begin{equation}
v_{e_{i,j}}^{(0)}= [h_{e_i} || h_{e_j}]
\end{equation}

\textbf{Node Embedding Update.} \ Then, we adopt relational graph transformer~\cite{DBLP:journals/ijon/BiCCLXZ24} to enhance the node features with the relational information from neighbor nodes. Each layer $l$ is similar to the transformer architecture. It takes a set of node embeddings $\textbf{V}^{(l)} \in \mathbb{R}^{N\times d_{in}}$ as input, and outputs a new set of node embeddings $\textbf{V}^{(l+1)} \in \mathbb{R}^{N\times d_{out}}$, where $N=|\mathcal{V}_e|+|\mathcal{V}_{ep}|$ is the number of nodes in LCG,  $d_{in}$ and $d_{out}$ are the dimensions of input and
output embeddings. 

In each layer, to integrate information from each neighbor, we adopt a shared self-attention mechanism~\cite{DBLP:conf/nips/VaswaniSPUJGKP17} to calculate the attention score:
\begin{gather}
\alpha_{ij} = \mathrm{softmax} (co_{ij})\\
co_{ij} = \frac{(v_i \textbf{W}_q)(v_j \textbf{W}_k)^\mathrm{T} }{\sqrt{d_k}}
\end{gather}
where $N_i$ is the first order neighbor set of node $i$, $co_{ij}$ measures the importance of neighbor $j$ to $i$,
$\textbf{W}_q, \textbf{W}_k \in \mathbb{R}^{d_{in}\times d_k}$ are learnable matrices, $d_k$ is a scaling factor to assign lower attention weights to uninformative nodes. 

Then we aggregate relational knowledge from the neighborhood information with weighted linear combination of the embeddings:
\begin{equation}
v_i^{(l+1)} = \sum_{j\in N_i} \alpha_{ij}^{(l)} (v_j^{(l)} \textbf{W}_v^{(l)})
\end{equation}
where $\textbf{W}_v^{(l)}\in \mathbb{R}^{d_{in}\times d_k}$ is a learnable matrix. We also adopt multi-head attention to attend to information from multiple attention heads. Thus, the output of the $l$-th layer for node $i$ is:
\begin{equation}
v_i^{(l+1)} = (\big|\big|_{c=1}^C \sum_{j\in N_i} \alpha_{ij}^{(l)} (v_j^{(l)} \textbf{W}_v^{(l)}))\textbf{W}_o^{(l)}
\end{equation}
where $C$ is the number of attention head and $\textbf{W}_o^{(l)}\in \mathbb{R}^{Cd_k \times d_{out}} $ is a learnable matrix.

\textbf{Measure Edge Heterogeneity.} \ \  It is intuitive that three types of edges in LCG contributes differently to ERE. Thus we propose to measure the edge heterogeneity and incorporate the edge features into node embeddings. Specifically, for each edge type in LCG, we learn a scalar:
\begin{equation}
\beta_t = r_t \textbf{W}_r
\end{equation}
where $1\leq t \leq T$, $T$ is the number of edge types, $r_t\in \mathbb{R}^{1\times d}$ denotes the edge features specific to the edge type, $\textbf{W}_r\in \mathbb{R}^{d\times 1}$ is a learnable matrix. $r_t$ will be randomly initialized. Then we incorporate $\beta_t$ as the attention bias into the attention score to adjust the interaction strength between two adjacent
nodes:
\begin{gather}
\widetilde{\alpha}_{ij} = \mathrm{softmax}  (\beta_t + co_{ij}) \label{eq.9}
\end{gather}

As the result, the final updated node embeddings considering the edge heterogeneity is:
\begin{equation}
\widetilde{v}_i^{(l+1)} = (\mathop{\big|\big|}\limits_{c=1}^C \sum_{j\in N_i} \widetilde{\alpha}_{ij}^{(l)} (v_j^{(l)} \textbf{W}_v^{(l)}))\textbf{W}_o^{(l)}
\end{equation}

By stacking multiple layers, the reasoning network could reach high-order interaction and maintain logic properties.

\textbf{Learning and Classification.}\ \ To predict whether there is the temporal or subevent relation between events $e_i$ and $e_j$, we concatenate the embeddings of "\texttt{[CLS]}", $e_i$, $e_j$ and the corresponding event pair as the logic enhanced representation. Thus, the probability distribution of the relation can be obtained through linear classification:
\begin{equation}
p_{e_{i,j}}=\mathrm{softmax} ([h_{\mathtt{[CLS]}} || \widetilde{v}_i || \widetilde{v}_j || \widetilde{v}_{i,j}] \textbf{W}_p)
\end{equation}
where $||$ denotes concatenation and $\textbf{W}_p$ is a learnable matrix.

For training, we adopt cross-entropy as the loss function:
\begin{equation}
\mathcal{L}_1= -\sum_{e_i,e_j \in \mathcal{E_D}} (1-y_{e_{i,j}}) \mathrm{log}(1-{p}_{e_{i,j}})+y_{e_{i,j}} \mathrm{log} ({p}_{e_{i,j}})
\end{equation}
where $y_{e_{i,j}}$ denotes the golden label.

\subsection{Joint Logic Learning}
Inspired by the logic-driven framework for consistency of \citet{DBLP:conf/emnlp/LiGMS19}, we further design two learning objectives by directly transforming the logical constraints into differentiable loss functions~\footnote{More details about joint logic learning can be found from appendix A: Details of Joint Logic Learning}.

\textbf{Symmetry Constraint.}\ \  Symmetry constraints indicate the event pair with flipping orders will have the reversed relation, the logical formula can be written as:
\begin{equation}
\mathop{\wedge}\limits_{e_i,e_j\in \mathcal{E_D}, r\in \mathcal{R}_{sym}} r(e_i, e_j) \leftrightarrow \bar{r}(e_j, e_i)
\end{equation}
where $\mathcal{R}_{sym}$ indicates the set of relations enforcing the
symmetry constraint. We use the product t-norm and transformation to the negative log space and obtain the symmetry loss:
\begin{equation}
\mathcal{L}_{sym}= \sum_{e_i,e_j\in \mathcal{E_D}} |\mathrm{log} (p_{e_{i,j}})-\mathrm{log} (\overline{p}_{e_{j,i}})|
\end{equation}

\textbf{Conjunction Constraint.}\ \ Conjunctive constraint are applicable to any three related events $e_i$, $e_j$ and $e_k$. It contributes to the joint learning of TRE and SRE. 
The conjunction constraint enforces the following logical formulas:
\begin{gather}
\mathop{\wedge}\limits_{e_i,e_j,e_k \in \mathcal{E_D}\atop {r_1, r_2}\in \mathcal{R}, r_3 \in De(r_1, r_2)}  r_1(e_i, e_j) \wedge r_2(e_j,e_k)  \rightarrow r_3(e_i, e_k) \\
\mathop{\wedge}\limits_{e_i,e_j,e_k \in \mathcal{E_D}\atop {r_1, r_2}\in \mathcal{R}, r_4 \notin De(r_1, r_2)}  r_1(e_i, e_j) \wedge r_2(e_j,e_k) \!\rightarrow \neg r_4(e_i, e_k) 
\end{gather}
where $De(r_1, r_2)$ is a set composed of all relations from $\mathcal{R}$ that do not conflict with $r_1$ and $r_2$.

Similarly, the loss function specific to conjunction constraint is:
\begin{gather}
\mathcal{L}_{conj}= \mathop{\sum}\limits_{e_i,e_j,e_k \in \mathcal{E_D}} |\mathcal{L}_{c_1}|+\mathop{\sum}\limits_{e_i,e_j,e_k \in \mathcal{E_D}}|\mathcal{L}_{c_2}|\\
\mathcal{L}_{c_1}= \mathrm{log} (p_{e_{i,j}})+\mathrm{log} (p_{e_{j,k}}) -\mathrm{log} (p_{e_{i,k}})\\
\mathcal{L}_{c_2}= \mathrm{log} (p_{e_{i,j}})+\mathrm{log} (p_{e_{j,k}}) -\mathrm{log} (1-p_{e_{i,k}})
\end{gather}

The final loss function combines the above logic learning and event relation learning objectives, where $\gamma$ are non-negative coefficients to control the
influence of each loss term:
\begin{equation}
\mathcal{L}= \mathcal{L}_1 + \gamma_{sym} \mathcal{L}_{sym} + \gamma_{conj} \mathcal{L}_{conj} \label{eq:20}
\end{equation}

\section{Experiments}
\subsection{Datasets and Metrics}
We evaluate logicERE on four widely used datasets. MATRES
~\cite{DBLP:conf/acl/RothWN18} and TCR
~\cite{DBLP:conf/acl/RothWNF18} are used to test the performance of TRE. HiEve
~\cite{DBLP:conf/lrec/GlavasSMK14} is used for SRE. MAVEN-ERE
~\cite{DBLP:conf/emnlp/WangC0PWL00LLLZ22} is used to test the joint learning performance. \textbf{MATRES} is a new dataset and mainly annotates four temporal relationships, i.e., BEFORE, AFTER, EQUAL and VAGUE. \textbf{TCR} is a small-scale dataset which has the same annotation scheme as MATRES. We only apply it to the testing phase. \textbf{HiEve} is a news corpus and annotates four subevent relationships, i.e., PARENT-CHILD, CHILD-PARENT, COREF and NOREL. \textbf{MAVEN-ERE} is a unified large-scale dataset that annotates data for event coreference, temporal, causal, and subevent relations. We only experiment on the event temporal and subevent relations. For TRE, it defines six relationships. To be consistent with our framework, we only consider type BEFORE and SIMULTANEOUS, and we manually annotate reflexive relationships AFTER and VAGUE, respectively. For SRE, it defines one relationships SUBEVENT and we manually annotate corresponding reflexive relationships SUPEREVENT. Note that HiEve and MAVEN-ERE provide ground-truth event coreference annotations, but MATRES does not. We follow \citet{DBLP:conf/acl/00010ZL23} and perform pre-training on MAVEN-ERE, and then use the pre-trained model to extract coreference data for MATRES. After the preprocessing steps, we add event-event edges $\mathcal{C}_{ee}$ to MATRES. For compatible comparison, we utilize the same data splits as in prior work for the considered datasets. We briefly summarize the data statistics for the above datasets in Table~\ref{tab:data}.

\begin{table}\small
	\centering
	\begin{tabular}{lc|c|c|c}
		Dataset & & Train &Dev & Test\\
		\midrule
		\multirow{2}{*}{MATRES}  &Document& 260& 21 &20          \\
		& Event pairs & 10,888 & 1,852& 840\\ 
		\midrule
		\multirow{2}{*}{TCR}  &Document& -& - &25         \\
		& Event pairs & - & -& 2,646\\ 
		\midrule
		\multirow{2}{*}{HiEve}  &Document& 80 & - &20         \\
		& Event pairs & 35,001 & -& 7,093\\ 
		\midrule
		{MAVEN-ERE}  &Document& 2,913  & 710  &857         \\
		(TRE)& Event pairs & 792,445 & 188,928& 234,844\\ 
		\midrule
		{MAVEN-ERE}  &Document& 2,913  & 710  &857         \\
		(SRE)& Event pairs & 9,193 & 2,826& 3,822\\ 
	\end{tabular}
	\caption{Data statistics for dataset MATRES, TCR, HiEve and MAVEN-ERE (TRE/SRE). }
	\label{tab:data}
\end{table}

We adopt the standard micro-averaged Precision (P), Recall (R) and F1-scores (F1) as evaluation metrics. All the results are the average of five trials of
different random seeds in each experiment.

\subsection{Parameter Settings}
Our implementation uses HuggingFace Transformers~\cite{DBLP:conf/emnlp/WolfDSCDMCRLFDS20} and PyTorch~\cite{DBLP:conf/nips/PaszkeGMLBCKLGA19}. We employ RoBERTa-base~\cite{DBLP:journals/corr/abs-1907-11692} as the document encoder. As for the input of the encoder, we set the dynamic window size to 256,
and divide documents into several overlapping windows with a step size 32. We use AdamW~\cite{DBLP:conf/iclr/LoshchilovH19} optimizer and learning rate is set to 2e-5. We adopt layer normalization~\cite{DBLP:journals/corr/BaKH16} and dropout~\cite{DBLP:journals/jmlr/SrivastavaHKSS14} between the high-order reasoning network layers. We perform early stopping and tune the hyper-parameters
by grid search on the development set: heads $C \in \{1,2,\mathbf{4},8\}$, dropout rate $\in \{0.1, \mathbf{0.2}, 0.3\}$ and loss coefficients $\gamma_{sym},\gamma_{conj} \in  \{0.1, \mathbf{0.2}, 0.4, 0.6\}$.

\subsection{Baselines}
We adopt the following state-of-the-art models for the MATRES, TCR, HiEve and MAVEN-ERE datasets respectively. 

For the MATRES dataset, we consider the following baselines. (1) \textbf{TEMPROB+ILP}~\cite{DBLP:conf/emnlp/NingSR19} incorporates temporal commonsense knowledge and integer linear programming; (2) \textbf{Hierarchical}~\cite{DBLP:journals/corr/abs-1904-08398} uses RoBERTa to encode different chunks of the document, and sets an additional BILSTM model to aggregate representations; (3) \textbf{Joint Constrain}~\cite{DBLP:conf/emnlp/WangCZR20} enforces consistency with joint constrained learning objectives; (4) \textbf{Self-Training}~\cite{DBLP:conf/emnlp/BallesterosAWPV20} relys on multitask and self-training techniques; (5) \textbf{Vanilla Classifier}~\cite{DBLP:conf/emnlp/WenJ21} is a common event relation classifier based on the RoBERTa; (6) \textbf{Relative Time}~\cite{DBLP:conf/emnlp/WenJ21} is a stack propagation framework employing relative time prediction as hints; (7) \textbf{Probabilistic Box}~\cite{DBLP:conf/acl/HwangLYPZM22} uses probabilistic boxes as implicit constraints to improve the consistency;  (8) \textbf{TGAGCN}~\cite{DBLP:conf/dasfaa/ZhuangHZ23} is based on event-specific syntactic dependency graph; (9) \textbf{OntoEnhance}~\cite{DBLP:journals/inffus/ZhuangFH23} fuses semantic information from event ontologies to enhance event representation;
(10) \textbf{SDLG}~\cite{DBLP:journals/ipm/ZhuangFH23} builds  syntax-based dynamic latent graph for event relation reasoning.

For TCR, the following baselines are included in our comparison. (1) \textbf{LSTM+knowledge}~\cite{DBLP:conf/emnlp/NingSR19} is a variant of TEMPROB+ILP that incorporates temporal commonsense knowledge in LSTM; (2) \textbf{HGRU}~\cite{DBLP:conf/emnlp/TanPH21} maps event embeddings to hyperbolic space; (3) \textbf{Poincaré Embeddings}~\cite{DBLP:conf/emnlp/TanPH21} learns rich event representations in hyperbolic spaces; (4) We also compare with \textbf{TEMPROB+ILP}, \textbf{Vanilla Classifier}, \textbf{TGAGCN}, \textbf{OntoEnhance} and \textbf{SDLG}.

For the HiEve dataset, we choose the following baselines. (1)\textbf{StructLR}~\cite{DBLP:conf/lrec/GlavasSMK14} is a supervised classifier combining event, bag-of-words, location, and syntactic features;  (2) \textbf{TACOLM}~\cite{DBLP:conf/acl/ZhouNKR20} uses duration prediction to assist SRE; (3) Similarly, we also compare with \textbf{Joint Constrain}, \textbf{Vanilla Classifier}, \textbf{Hierarchical},  \textbf{TGAGCN}, \textbf{OntoEnhance} and \textbf{SDLG}. 

For MAVEN-ERE, we select RoBERTa~\cite{DBLP:journals/corr/abs-1907-11692} as the main baseline, which adopts RoBERTa as the document encoder and obtains event embeddings for pair-wise classification. According to whether multiple event relations are trained simultaneously, we consider two settings, i.e., (1) \textbf{RoBERTa$_{split}$} and (2) \textbf{RoBERTa$_{joint}$} for comparisons. We also compare with (3) \textbf{GraphERE$_{split}$} and (4) \textbf{GraphERE$_{joint}$}~\cite{DBLP:journals/corr/abs-2403-12523}, which adopt graph-enhanced event embeddings for pair-wise classification. Besides, for each tasks, we explore related baselines in our experiments: (5) \textbf{DocTime}~\cite{DBLP:conf/naacl/MathurMKGDTNMJ22} constructs temporal dependency graph for TRE; (6) \textbf{MultiFeatures}~\cite{DBLP:conf/acl/AldawsariF19} employs discourse and narrative features for SRE.

\subsection{Comparison}

Table~\ref{tab:MATRES},~\ref{tab:TCR},~\ref{tab:HiEve} and~\ref{tab:MAVEN-ERE} show the performance of the models on four datasets. Here, the performance for the models in previous work is inherited from the original papers. 

\begin{table}
	\centering
	\begin{tabular}{l|c|c|c}
		Model &  P & R & F1\\
		\midrule
		TEMPROB+ILP (2019) & 71.3  & 82.1 &76.3          \\
		Hierarchical (2019) & 74.2 & 83.1 & 78.4       \\
		Joint Constrain (2020)& 73.4 &85.0&78.3\\ 
		Self-Training (2020) & -& - &81.6         \\
		Vanilla Classifier (2021)& 78.1 &82.5& 80.2\\ 
		Relative Time (2021) &78.4 & 85.2  &81.7  \\
		Probabilistic Box (2022)& - & -& 77.1\\ 
		TGAGCN (2023) & 81.1 & 83.0 & 82.0\\ 
		OntoEnhance (2023)  & 79.0 & 86.5 & 82.6 \\
		SDLG (2023) & 82.0 &84.2 & 83.1    \\
		\midrule
		LogicERE (ours)  & 82.9  & 84.5 & \textbf{83.7}\\
	\end{tabular}
	\caption{Model performance on test data of MATRES for temporal relation extraction.}
	\label{tab:MATRES}
\end{table}

\begin{table}[t]
	\centering
	\begin{tabular}{l|c|c|c}
		Model &  P & R & F1\\
		\midrule
		LSTM+knowledge (2019) & 79.3 & 76.9  &78.1        \\
		TEMPROB+ILP (2019) & -  & - & 78.6         \\
		Vanilla Classifier (2021)& 89.2&76.7 & 82.5\\ 
		HGRU (2021) &88.3 &79.0 & 83.5     \\
		Poincaré Embeddings (2021)& 85.0 &86.0& 85.5\\ 
		TGAGCN (2023) & 89.2 & 84.3 & 86.7\\ 
		OntoEnhance (2023)  & 89.6 & 84.3 & 86.8 \\
		SDLG (2023) & 88.3 & 87.0& 87.6   \\
		\midrule
		LogicERE (ours) & 90.8  & 86.7 & \textbf{88.7} \\
	\end{tabular}
	\caption{Model performance on test set of TCR for temporal relation extraction.}
	\label{tab:TCR}
\end{table}

\begin{table}[t]
	\centering
	\begin{tabular}{l|c|c|c}
		\multirow{2}{*} {Model} &\multicolumn{3}{c} {F1} \\
		\cmidrule{2-4}
		&  PC & CP & Avg.\\
		\midrule
		StructLR (2014) & 52.2 & 63.4  & 57.7       \\
		Hierarchical (2019) & 63.7 & 57.1 & 60.4      \\
		TACOLM (2020) & 48.5 & 49.4 & 48.9        \\
		Joint Constrain (2020)& 62.5 & 56.4& 59.5\\ 
		Vanilla Classifier (2021)& 62.8  & 52.3 & 57.5\\ 
		TGAGCN (2023) & 65.2  & 56.0  & 60.6\\ 
		OntoEnhance (2023)  & 66.7  & 57.3  & 62.0\\
		SDLG (2023) & 64.9  & 58.5  & 61.7  \\
		\midrule
		LogicERE (ours) & 68.3  & 62.5 & \textbf{65.4}\\
	\end{tabular}
	\caption{Model performance on test set of HiEve for subevent relation extraction. We focus on the performance for PARENT-CHILD (PC), CHILD-PARENT (CP), and their micro-average (Avg.).}
	\label{tab:HiEve}
\end{table}

\textbf{TRE Evaluation.}\ \ From the results in Table~\ref{tab:MATRES} and~\ref{tab:TCR} for temporal relation extraction, we can draw the following observations: 1) LogicERE significantly outperforms previous models (with $p\  \textless\  0.05$) on both datasets, which
demonstrates its effectiveness for TRE. 2) Compared with other methods considering logic coherence (i.e., Joint Constrain and Probabilistic Box), LogicERE gains at least $5.4\%$ F1 scores improvements on MATRES. This indicates that LogicERE is more effective in incorporating logic constraints in graph-enhanced event embeddings, which preserves coherence of temporal relations and enriches the features in event embeddings. 3) LogicERE surpasses the graph-based model (i.e., TGAGCN, OntoEnhance and SDLG) by at least $0.6\%$ and $1.1\%$ F1 scores improvements on MATRES and TCR respectively. It shows the advantages of logic constraint induced graph for TRE.  Importantly, LogicERE can accomplish state-of-the-art performance for TRE without any external knowledge. This
is different from recent work that requires additional resources to secure good performance, such as ontology knowledge in OntoEnhance or dependency parsing in TGAGCN and SDLG.

\begin{table*}[t]
	\centering
	\begin{tabular}{l|c|c|c|c|c|c}
		\multirow{2}{*}{Model}&\multicolumn{3}{c|}{TRE}&\multicolumn{3}{c}{SRE}\\
		\cmidrule{2-4} \cmidrule{5-7}
		& P & R & F1 & P & R & F1\\
		\midrule
		DocTime (2022) &54.4 &53.0 &53.7  &- &- &- \\
		MultiFeatures (2019) &- &- &-  &16.4 &19.9 &18.0\\
		\midrule
		RoBERTa$_{split}$ (2024)&53.1 &53.3 &53.2  &27.3 &20.9 &23.7 \\
		RoBERTa$_{joint}$ (2024)&50.7&56.7& 53.5 & 19.9 &24.2&21.8 \\
		GraphERE$_{split}$ (2024)&57.1&50.0&53.3 &26.1&20.8&23.1 \\
		GraphERE$_{joint}$ (2024) &55.1&54.4&54.7 &31.2 &24.2 &27.3  \\
		\midrule
		LogicERE$_{split}$ (ours)
		& 58.2 &50.5 & 54.1 & 29.4 &22.8 & 25.7  \\
		LogicERE$_{joint}$ (ours)
		& 56.3 & 53.7 & \textbf{55.0} & 32.8 & 28.2& \textbf{30.3} \\
	\end{tabular}
	\caption{Model performance on test set of MAVEN-ERE for  temporal relation extraction and subevent relation extraction, split/joint stands for separately or jointly training TRE and SRE tasks.}
	\label{tab:MAVEN-ERE}
\end{table*}

\begin{table}[t]
	\centering
	\begin{tabular}{l|c|c}
		Model &  TRE & SRE \\
		\midrule
		LogicERE$_{joint}$ (full) & \textbf{55.0} & \textbf{30.3}        \\
		\midrule
		w/o edge heterogeneity & 54.4 {\scriptsize (-0.6)} & 29.5 {\scriptsize (-0.8)}      \\
		w/o coreference& 54.7 {\scriptsize (-0.3)} & 30.0 {\scriptsize (-0.3)}\\ 
		w/o event-event pair edges & 53.5 {\scriptsize (-1.5)}& 28.8 {\scriptsize (-1.5)}       \\
		w/o symmetry objective& 54.8 {\scriptsize (-0.2)}& 30.0 {\scriptsize (-0.3)}\\ 
		w/o conjunction objective & 54.5 {\scriptsize (-0.5)}  & 29.7 {\scriptsize (-0.6)}\\
		w/o joint logic learning & 54.1 {\scriptsize (-0.9)} & 28.9 {\scriptsize (-1.4)}\\
	\end{tabular}
	\caption{Ablation results (F1 scores) on MAVEN-ERE.}
	\label{tab:ablation}
\end{table}

\textbf{SRE Evaluation.}\ \ The results in Table~\ref{tab:HiEve} for subevent relation extraction exhibit similar observation: 1) LogicERE can effectively deal with SRE task, achieving a significant advantage over all methods (with $p\  \textless\  0.05$). The reason may be that our model can benefit from high-order reasoning of subevent relations through logic induced reasoning network and can get logic-enhanced event embeddings for SRE. 2) Consistent with experimental results on MATRES and TCR, LogicERE still shows significant advantages over other logic-enhanced models (i.e., Joint Constrain) and graph-based models (i.e., TGAGCN, OntoEnhance and SDLG). This verifies the effectiveness of our model for SRE.  

\textbf{Joint Learning Evaluation.}\ \  MAVEN-ERE contains annotations for both TRE and SRE. We experiment on it to evaluate the ability of LogicERE in jointly learning multiple event relations. From Table~\ref{tab:MAVEN-ERE}, we observe that: 1) The proposed LogicERE$_{joint}$ surpasses all baselines in F1 scores for both event relations. Compared with the best existing methods for each task, our model improves by $0.3\%$ in TRE and $3.0\%$ in SRE. 2) LogicERE has a considerable improvement in Precision. LogicERE$_{joint}$ gains on average $1.4\%$ Precision improvement on the two tasks (LogicERE$_{split}$ for $1.6\%$). This indicates that we incorporate logic constraints into LCG which enriches the features in event embeddings through high-order reasoning, and improves the performance in Precision. 3) The models in joint setting outperform corresponding ones in split setting. Meanwhile, the improvement from LogicERE$_{split}$ to LogicERE$_{joint}$ are $0.9\%$ and $4.6\%$ on TRE and SRE respectively, indicting that learning from other types of relations boost the ERE performance. 4) SRE is a challenging task with almost all the models exhibiting less than $30\%$ in F1 Score. However, LogicERE$_{joint}$ has the most significant advantage of
$3.0\%$ in F1 Score over baselines. The possible reason is that the global
consistence ensured by LCG and joint logic learning naturally makes up for the weak supervision signals for SRE. 

In general, the experimental results show that LogicERE can effectively combine the supervision signals from two tasks, assisting in the comprehension of both temporal relation and subevent relation. 

\subsection{Ablation Study}
We then conduct ablation study to elucidate the effectiveness of main components of our model. Likewise, we only present the results
on MAVEN-ERE. In particular, we consider the following internal
baselines. (1) 
\textbf{w/o edge heterogeneity} does not consider the edge heterogeneity, and thus the scalar $\beta_t$ is removed from Eq.~\ref{eq.9} in the relational knowledge aggregation process. 
(2) \textbf{w/o coreference} removes  $\mathcal{C}_{ee}$ from LCG and do not use ground-truth coreference annotations as training labels. (3) \textbf{w/o event-event pair edges}  removes $\mathcal{C}_{ep}$ from LCG. (4) \textbf{w/o symmetry objective}  removes $\mathcal{L}_{sym}$  from Eq.~\ref{eq:20}. (5) \textbf{w/o conjunction objective}  removes $\mathcal{L}_{conj}$  from Eq.~\ref{eq:20}. (6) \textbf{w/o joint logic learning} removes the joint logic learning objective and adopt $\mathcal{L}_1$ as the loss function.

Results are shown in Table~\ref{tab:ablation}. We can observe that: 1) Our full model significantly outperforms all internal baselines on MAVEN-ERE.
Compared to LogicERE, w/o. edge heterogeneity drops $0.6\%$ and $0.8\%$ F1 scores for TRE and SRE respectively, validating the necessity of capturing the semantic information of different types of edges in LCG. 2) w/o. event-event pair edges shows huge performance drop. This result demonstrates that the deep information interaction between the event and the
event pair contributes to more informative event embeddings, impelling better event pair relation inference. 3) Experimental results of w/o symmetry objective, w/o conjunction objective and w/o joint logic learning demonstrate that, after removing either of the two, the F1 scores go down. It indicates that these two kinds of logic constraints both contribute to our model. Simultaneously using two kinds of knowledge further improves the overall performance.

\section{Conclusion}
We present a novel logic induced high-order reasoning network (LogicERE) to enhance the event relation reasoning with logic constraints. We first design a logic constraint induced graph (LCG) which contains interactions between events and among event pairs. Then we encode the heterogeneous LCG for high-order event relation reasoning while maintaining inherent logic properties of event relations. Finally, we further incorporate logic constraints with joint logic learning. Extensive experiments show that LogicERE can effectively utilize logic properties to enhance the event and event pair embeddings, and achieve state-of-the-art performance for both TRE and SRE. The joint learning evaluation reveals that LogicERE can effectively maintain global consistency of two types relations, assisting in the comprehension of both temporal and subevent relation.

\section{Acknowledgments}
We thank all the anonymous reviewers and meta reviewers for their valuable comments, as well as all of our team members for their support and assistance. This work was partially supported by National Key R\&D Program of China No. 2022YFB3103600, NSFC (Nos. U23A20296, 72371245, 62476283, 62376284).

\label{sec:reference_examples}
\nobibliography*
\bibliography{aaai25}

\appendix
\section{A. Details of Joint Logic Learning}

\begin{figure*}[t]
	\centering
	\includegraphics[width=1\linewidth]{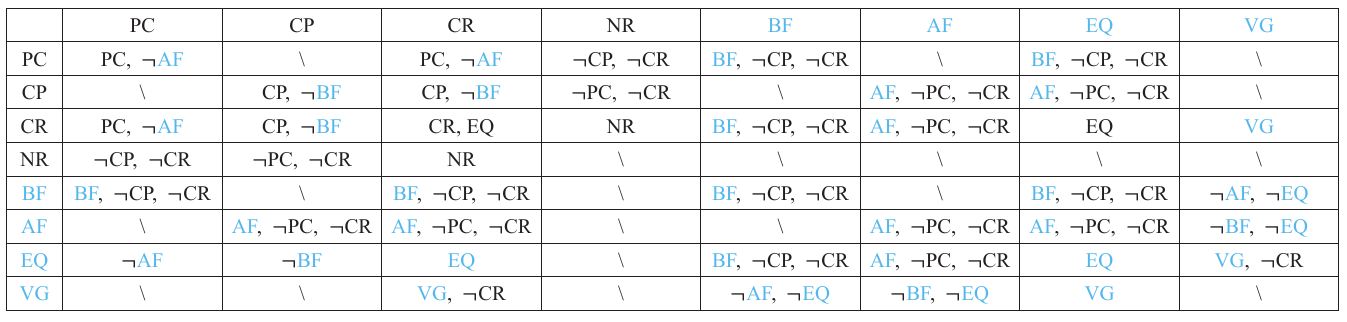}
	\caption{The induction table for conjunctive constraints on temporal and subevent relations. The abbreviations PC, CP , CR, NR, BF, AF, EQ and VG denote PARENT-CHILD, CHILD-PARENT, COREF, NOREL, BEFORE, AFTER, EQUAL and VAGUE, respectively. Subevent relations are in black, and temporal relations are in blue. ``$\backslash$'' denotes no constraints.}
	\label{1}
\end{figure*} 

With the logical formulas of corresponding logic constraints, we now  focus on the unification of discrete declarative logic constraints with the loss-driven learning paradigm. To address this, we use relaxations of logic in the form of t-norms to deterministically compile rules into differentiable loss functions. Different t-norms map the Boolean operations into different continuous functions.

We use the product t-norm as it strictly generalizes the widely
used cross entropy loss. The mapping of standard Boolean
operations into continuous functions for product t-norm is shown in Table~\ref{tab:map}.

\begin{table}
	\centering
	\begin{tabular}{l|c|c}
		Name &  Boolean Logic  & Product \\
		\midrule
		Negation  & $\neg A $ & $1-a$        \\
		T-norm & $A\land B$ & $ab$     \\
		T-conorm & $A\lor B$  & $a+b-ab$\\ 
		Residuum & $A \rightarrow B$& $\min (1, b/a)$  \\
	\end{tabular}
	\caption{Mapping discrete statements into differentiable functions. Literals are upper-cased, and real-valued probabilities are lower-cased.}
	\label{tab:map}
\end{table}

For the \textbf{symmetry constraint}, we have:
\begin{equation}
\mathop{\wedge}\limits_{e_i,e_j\in \mathcal{E_D}, r\in \mathcal{R}_{sym}} r(e_i, e_j) \leftrightarrow \bar{r}(e_j, e_i)
\end{equation}
where $\mathcal{R}_{sym}$ indicates the set of relations enforcing the
symmetry constraint. 

Using the product t-norm, we get:
\begin{equation}
\prod_{e_i,e_j\in \mathcal{E_D}, r\in \mathcal{R}_{sym}}\!\!\!\!\!\!\!\!\! \min (1, \bar{r}_{(e_j, e_i)}/r_{(e_i, e_j)} )\!\! \min (1, r_{(e_i, e_j)} / \bar{r}_{(e_j, e_i)})
\end{equation}

Transforming to the negative log space, we get the symmetry loss:
\begin{equation}
\mathcal{L}_{sym}= \sum_{e_i,e_j\in \mathcal{E_D}} |\mathrm{log} r_{(e_i, e_j)} -\mathrm{log} \bar{r}_{(e_j, e_i)}|
\end{equation}

For the \textbf{conjunction constraint}, the induction table for conjunctive constraints on temporal and subevent relations is shown in Figure~\ref{1}.

we have:
\begin{gather}
\mathop{\wedge}\limits_{e_i,e_j,e_k \in \mathcal{E_D}\atop {r_1, r_2}\in \mathcal{R}, r_3 \in De(r_1, r_2)}  r_1(e_i, e_j) \wedge r_2(e_j,e_k)  \rightarrow r_3(e_i, e_k) \\
\mathop{\wedge}\limits_{e_i,e_j,e_k \in \mathcal{E_D}\atop {r_1, r_2}\in \mathcal{R}, r_4 \notin De(r_1, r_2)}  r_1(e_i, e_j) \wedge r_2(e_j,e_k) \!\rightarrow \neg r_4(e_i, e_k) 
\end{gather}
where $De(r_1, r_2)$ is a set composed of all relations from $\mathcal{R}$ that do not conflict with $r_1$ and $r_2$.

Similarly, we get the loss function for the conjunction constraint:
\begin{gather}
\mathcal{L}_{conj}= \mathop{\sum}\limits_{e_i,e_j,e_k \in \mathcal{E_D}} |\mathcal{L}_{c_1}|+\mathop{\sum}\limits_{e_i,e_j,e_k \in \mathcal{E_D}}|\mathcal{L}_{c_2}|\\
\mathcal{L}_{c_1}= \mathrm{log} (p_{e_{i,j}})+\mathrm{log} (p_{e_{j,k}}) -\mathrm{log} (p_{e_{i,k}})\\
\mathcal{L}_{c_2}= \mathrm{log} (p_{e_{i,j}})+\mathrm{log} (p_{e_{j,k}}) -\mathrm{log} (1-p_{e_{i,k}})
\end{gather}

\end{document}